\documentclass[twocolumn,prl,a4paper,superscriptaddress]{revtex4}
\usepackage{comment}
\usepackage{color}
\usepackage{amsmath}
\usepackage{amssymb}
\usepackage{graphicx}
\usepackage{braket} 
\usepackage{esint}
\usepackage{adjustbox} 
\usepackage[utf8]{inputenc}

\usepackage[unicode=true,pdfusetitle,
bookmarks=true,bookmarksnumbered=false,bookmarksopen=false,
breaklinks=true,pdfborder={0 0 1},backref=false,colorlinks=true]
{hyperref}
\hypersetup{
	linkcolor=red,urlcolor=blue,citecolor=blue,anchorcolor=blue} 
\usepackage{ulem} 

\usepackage{dcolumn}\usepackage{bm}

\makeatother 
\begin{document}

\title{Thermal Entanglement of superconducting qubits for arbitrary interaction strength} 


\author{Areeda Ayoub}   
\author{Javed Akram}
 
\email{javedakram@daad-alumni.de}
 
\affiliation{Department of Physics, COMSATS University Islamabad, 44000, Islamabad Pakistan}
\date{\today}
\begin{abstract}

We investigate the thermal entanglement in two superconducting qubits for arbitrary interaction strength and ground state frequencies. We calculate the concurrence of the system to quantify the thermal entanglement. We suggest a scheme, where an external tunable coupler qubit sandwich between two superconducting qubits generates  entanglement. The behavior of concurrence is analyzed for three different cases, in which we consider the effects of the temperature, the qubit-qubit effective coupling strength, and the qubit frequencies on the thermal entanglement. What deserves mentioning here is that to achieve maximally entangled states, it is better to use two superconducting qubits with the same frequencies. We also note that for a given temperature, the thermal entanglement can be tuned by qubit internal capacitance and inductance.   
\end{abstract}

\maketitle

\section{Introduction}  \label{Sec-1}
From the first idea of a quantum computer by Feynman in $1982$, the race for the transition from classical computers to the quantum computer has been going on. The key to achieve this is by replacing the classical bit with the superconducting qubit. Superconducting qubits have attracted the attention of scientists not only for their high speed and efficiency but the versatile nature of their quantum states that can be engineered and controlled \cite{Ventura-2000,Akram-2008,Ladd-2010,Kelly-2015}. A physical system that is required to process quantum information must obey quantum mechanical laws like superposition and entanglement. The states of the composite system are entangled when they cannot be defined in the form of the product of states of the individual system.  The simplest composite system that can exhibit entanglement consists of two-level systems (with ground $\ket{0}$ and excited $\ket{1}$ states). To perform quantum information processing, superconducting qubits are the most suitable candidates. Therefore, the generation of superconducting qubits and processing of information with these qubits are the main questions that need to be investigated. Additionally, it is also equally important to explore environmental noises and temperature effects on the superconducting qubits \cite{PhysRevLett.119.180511,Bernien-2017,Zhang-2017,Sung_2020,Xiang_Ping_2006}.  The coherence of superconducting qubits is greatly improved by reducing their sensitivity to the charge noise  via   shunting a large capacitance to the Josephson junction. This was first proposed for capacitively shunted flux qubit in \cite{PhysRevB.75.140515} and later experimentally realized in \cite{PhysRevLett.105.100502,Yan-2016}. Later it was also proposed for the Transmon qubits in \cite{PhysRevA.76.042319} and implemented in \cite{PhysRevLett.111.080502}. Nowadays, the main focus of scientific research is to preserve these states from decoherence by taking advantage of effective coupling,  which can be constructed between the superconducting qubits by interaction with a common tunable coupler \cite{An-2007}.

From the last few years, entanglement has been thoroughly investigated \cite{book:1162666,Chruscinski-2011}. Entanglement has non-classical nature, therefore it helps to study conceptual foundations of quantum mechanics. Interpretation of quantum information processing is not possible without an understanding of quantum entanglement, and interaction of entanglement with the environment becomes very sensitive to do quantum computing and quantum information processing. Experimentalists always want to preserve entanglement for a longer period. However, in the real world, it is quite tough to protect the system from the environment, especially from the thermal environment. Therefore, in this paper, we investigate the effect of temperature on the entanglement, which we generated by using two coupled superconducting qubits \cite{PhysRevA.72.034302,PhysRevLett.80.2245,PhysRevLett.87.017901}. Additionally, we also study the effect of  interaction strength and ground state frequencies on the thermal entanglement.          

In this work, we introduce the simple and widely applicable method of using a tunable coupler to couple two superconducting qubits to generate entanglement. It consists of generic three-body systems with exchange-type interaction. The coupler is a central component that not only helps to exchange interaction between two qubits but also helps to tune the coupling strength between the next-nearest neighbor 
\cite{PhysRevLett.90.127901}.

For a quantum mechanical system  in thermal equilibrium at temperature $T$, the density matrix can be composed as $\rho=e^{-\beta H}/Z$, where, $H$ describes the Hamiltonian of the system, $Z=tr e^{-\beta H}$ defines the partition function and $\beta={1}/({k_B T})$ where $k_B$ illustrate the Boltzman constant. The thermal entanglement of two qubit system can be determined by the concurrence $C_E$ which is given by the \textit{Wootter's} formula \cite{PhysRevLett.78.5022} as $C_E=max[0,2~max[\lambda_i]-\sum_{i=1}^{4}\lambda_i]$. Here $\lambda_i$ is the square root of the eigen values of the matrix $R=\rho(\sigma_{1}^{y}\otimes\sigma_{2}^{y})\rho^{*}(\sigma_{1}^{y}\otimes\sigma_{2}^{y})$, where $*$ represents the complex conjugate. The concurrence $C_E$ can be defined for both mixed and pure states, ranges from $0$ to $1$. In general, we can define the two qubit composite system state as
 
\begin{figure} 
    \centering
    \includegraphics[height=4cm,width=8cm]{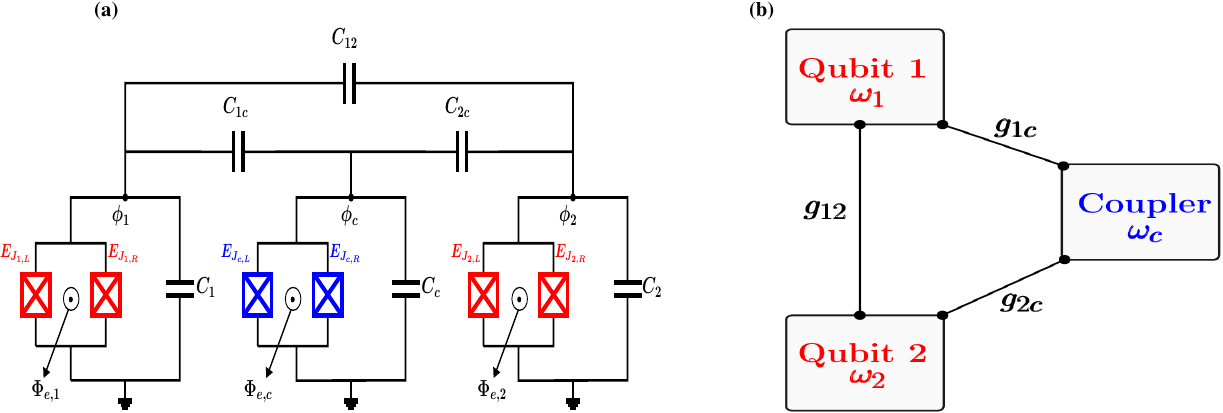}
    \caption{Schematic diagram of implementing superconducting qubits, consisting of qubit modes $(1,2)$   and coupler mode $c$. Here,  $C_\lambda$ is the dominant mode capacitance, $C_{jc}$ describes the coupling capacitance between qubit $j$  and coupler, $C_{12}$ illustrates the direct coupling capacitance between the qubits and $\phi_\lambda$ represents the reduced node flux. Here, $E_{J_{\lambda,L(R)}}$ is the defined as a Josephson energy of the left (right) junction mode $\lambda=(1,2,c)$.  In Fig. (b) we plot a block diagram of two qubits coupled with tunable coupler $\omega_c$, where qubit have frequencies $\omega_1$ and $\omega_2$}
    \label{Figure01}
\end{figure} 

\begin{equation}
    \ket{\phi}=a\ket{00}+b\ket{01}+c\ket{10}+d \ket{11},
\end{equation}\label{1}
here, $\ket{\phi}\in H_A\otimes H_B$, the two systems are separable if and only if $ad=bc$. Therefore, we can define concurrence as $ C_{E}(\psi)=2|ad-bc|.$ 
In this respect, we organize this paper as follows. We discuss the modeling of the qubit and coupled qubits in Sec. \ref{Sec-2}.  We introduce a method in Sec. \ref{Sec-3}, through which interaction between the superconducting qubits can be engineered to implement two coupled qubit entangling operations. We also discuss how temperature, effective qubit-qubit coupling strength, and qubit frequencies affect thermal entanglement. The conclusion is given in Sec. \ref{Sec-4}.

\section{Theoretical Model} \label{Sec-2}
\subsection{Modeling of a Qubit}
 A superconducting circuit needs some non-linear element to function as a qubit and this non-linearity is provided by the Josephson Junction, which forms the backbone of superconducting circuits. Josephson junction is based on two superconductors separated by a very thin layer of an insulator to allow tunneling of discrete charges across the junction \cite{You-2011}, the superconducting qubit circuit consists of a capacitor and Josephson junction as shown in Fig. \ref{Figure01}(a). The energy of the circuit elements is defined as 
\begin{equation}\label{3}
     E(t)=  \int_{-\propto}^{t}V(t')I(t')dt'.
\end{equation}
where $V(t')$ and $I(t')$ represent the voltage and current of the capacitor or non-linear inductor. According to Faraday's law, flux can be defined as $\Phi(t)=  \int_{-\propto}^{t}V(t')dt'.$ By using relation for the current flowing through the capacitor $I=C\frac{dV}{dt}$, we derive
the kinetic energy stored in the capacitor with capacitance $C$ in terms of  node flux as $T=\frac{1}{2}{\Big(\frac{\hbar}{2e}\Big)}^{2}C\Dot{\phi}^{2}$,
where $\phi=2\pi\Phi/{\Phi_0}$ is the reduced flux and ${\Phi_0}=h/2e$ is the superconducting magnetic flux. The potential energy of the Josephson junction can be derived by using Eq. (\ref{3}) as $ U=E_J(1-\cos{\Phi}),$
where $E_J=\hbar I_c/2e$ is the Josephson energy which  depends on the barrier transparency and superconducting gap \cite{Wendin-2007}. The Hamiltonian of the single qubit can be written as
\begin{equation}\label{4}
    \hat{H}=4E_C {\hat{n}}^2-E_J(1-\cos{\hat{\phi}}),
\end{equation}
where $\hat{n}=q/2e$ is the number of Cooper pairs and $E_C=e^{2}/2C$ defines the charging energy of the capacitor. Since $\hat{\phi}$ is very small, by expanding the potential term of the Eq. (\ref{4}) into power series, we get approximate Hamiltonian as
\begin{equation}\label{5}
    \hat{H}=4E_C {\hat{n}}^2+\frac{1}{2!}E_J{\hat{\phi}}^{2}-\frac{1}{4!}E_J{\hat{\phi}}^{4}.
\end{equation}
By introducing the creation $(\hat{b}^{\dag})$  and annihilation $(\hat{b})$ above Hamiltonian can be rewritten as
\begin{equation}\label{6}
 {H}=\sqrt{8E_JE_C}\Big(\hat{b}^\dag \hat{b}+\frac{1}{2} \Big)-\frac{E_C}{12}{|\hat{b}^\dag+\hat{b}|^4},
\end{equation}
where $\hat{n}$ and $\hat{\phi}$ are defined as $\frac{\Dot\iota}{2}\sqrt[^4]{\frac{E_J}{2E_C}} (\hat{b}^\dag-\hat{b})$ and $\sqrt[^4]{\frac{2E_C}{E_J}}(\hat{b}^\dag+\hat{b}),$ respectively.
Here, annihilation and creation operators follow commutation relation as $ [\hat{b},\hat{b}^{\dag}]=1.$
 Neglecting the constant term and fast-oscillating terms, the Hamiltonian of the system resembles the Duffing oscillator ($\hbar=1$), which is a harmonic oscillator plus an additional quartic term
\begin{equation}\label{7}
 {H}=\omega \hat{b}^\dag\hat{b}+\frac{\alpha}{2}{\hat{b}^\dag \hat{b}^\dag\hat{b} \hat{b}}.
\end{equation}
 where $\omega=\sqrt{8E_JE_C}- E_C$  defines the transition frequency and $\alpha=-E_C$  describes the anharmonicity in the system. If the anharmonicity is sufficiently large then excitation to higher states is suppressed and the lowest two levels can be treated as a two-level qubit system. And the quantum two-level simplified Hamiltonian can be represented as
\begin{equation}
    H=\omega\frac{\sigma_z}{2},
\end{equation}
where $\sigma_z$ is the Pauli z-operator.  It is necessary to keep in mind that higher states exist, however, their influence on the system dynamics is quite small.

\begin{figure}  
    \includegraphics[height=7cm,width=8cm]{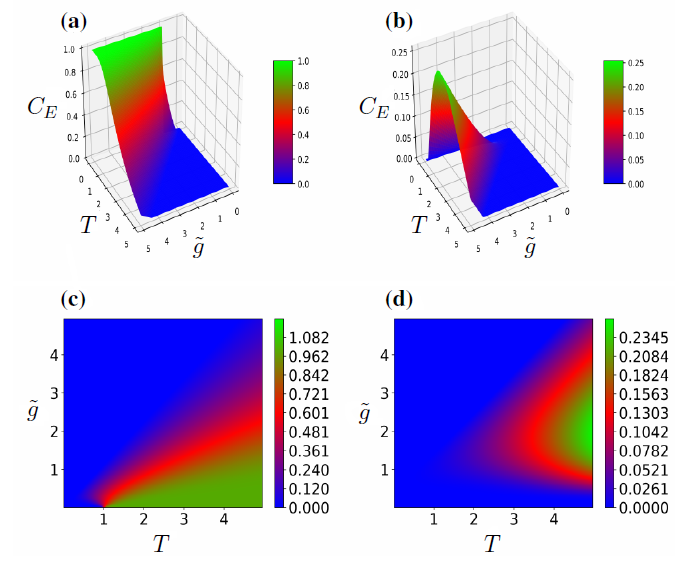} 
    \caption{The two-dimensional concurrence is plotted versus temperature $T$  and qubit-qubit effective coupling strength $\tilde{g}$ for two coupled superconducting qubits. The left panel corresponds to   $\tilde{\omega}_1=\tilde{\omega}_2=1$ case and right panel corresponds to $\tilde{\omega}_1=5$ and $\tilde{\omega}_2=1$ case. $T$ is plotted in units of Boltzmann's constant $k_B=1$ and we work in units where $\tilde{g}$ and $\tilde{\omega}_j$ are dimensionless.}
    \label{Figure02}
\end{figure}

\subsection{Modeling of Coupled Qubits}
In this subsection, we discuss the coupling of superconducting qubits with a tunable coupler, as shown in Fig. \ref{Figure01}(a). Each qubit is treated as a weakly anharmonic oscillator consisting of capacitor $C_\lambda$ and  Josephson junction, where the subscript defines as $\lambda=\{1,2,c\}$. For simplicity, the junction capacitance is merged into  $C_\lambda$. The kinetic and potential energies of these coupled qubit systems are given as
\begin{equation}
\begin{split}
     T=&  \frac{1}{2}\Big[C_1{\Dot{\Phi_1}}^{2}+C_c\Dot{\Phi_c}^{2}+C_2\Dot{\Phi_2}^{2}+C_{1c}(\Dot{\Phi_1}-\dot{\Phi_c})^{2}\\
     &+ C_{2c}(\Dot{\Phi_2}-\Dot{\Phi_c})^{2}+C_{12}(\Dot{\Phi_1}-\Dot{\Phi_2})^{2}\Big],
     \end{split}\label{8}
\end{equation}
\begin{equation}           
\begin{split}
       U=&{E_{J_1}}\Big[1-\cos{\big(\frac{2\pi}{\Phi_0}\Phi_1\Big)}\Big]+{E_{J_c}}\Big[ 1-\cos{\big(\frac{2\pi}{\Phi_0}\Phi_c\Big)}\Big]\\
       & +{E_{J_2}}\Big[1-\cos{\big(\frac{2\pi}{\Phi_0}\Phi_2\Big)}\Big],
\end{split} \label{9} 
\end{equation}
where $T$ labels the kinetic energy and $U$ specifies the potential energy.  Here, Josephson Energy of individual qubits is defined as a combination of left and right Josephson junction individual energies,  
 
\begin{equation}\label{9-a}
     E_{J_{\lambda}}= E_{J_{\lambda,T}}\sqrt{\cos^2\big(\frac{\pi \Phi_{e,\lambda}}{\Phi_{0}}  \big) + d^2_{\lambda} \sin^2\big(\frac{\pi \Phi_{e,\lambda}}{\Phi_{0}}  \big)}, 
 \end{equation}
 
 where $\Phi_{0} = h/2e$ is described as a superconducting quantum flux, here $E_{J_{\lambda,T}}= E_{J_{\lambda,L}}+E_{J_{\lambda,R}}$ is represented as a sum of the Josephson energies, and $d_{\lambda}= \big(E_{J_{\lambda,L}}-E_{J_{\lambda,R}}\big)/\big(E_{J_{\lambda,L}}+ E_{J_{\lambda,R}} \big)$ is the junction asymmetry \cite{PhysRevA.76.042319}.  
In compact form, the kinetic energy of the coupled two-qubit system can be defined as $T=\frac{1}{2}\Dot{\Vec{\Phi}}^{T}\textbf{C}\Dot{\Vec{\Phi}}$, here $\textbf{C}$ describes a 3$\times$3 capacitance matrix \cite{PhysRevApplied.10.054062},
 \begin{equation}\label{10}
     \textbf{C} = \begin{pmatrix} 
   C_{112}+C_{1c}&  -C_{1c} & -C_{12} \\\\
  -C_{1c} &  C_{c2c}+C_{1c}  & -C_{2c}\\\\ -C_{12}&-C_{2c}&C_{212}+ C_{2c}
    \end{pmatrix},
 \end{equation}
where, $C_{112}=C_{1}+C_{12}$, $C_{c2c}=C_{c}+C_{2c}$ and $C_{212}=C_{2}+C_{12}$. The Hamiltonian of the system can be depicted as
\begin{equation}\label{11}
    H=\sum_{\lambda}q_{\lambda}\Dot{\Phi}_{\lambda}-\mathcal{L}=\frac{1}{2}{\Vec{q}}^{T}[\textbf{C}^{-1}]\Vec{q}+U,
\end{equation}
here, $\textbf{C}^{-1}$ describes the inverse of the  capacitance matrix and Cooper-pair number operator is defined as $\hat{n}_{\lambda}=\hat{q}_{\lambda}/2e$. The total  Hamiltonian of this coupled  system can be disclosed as
 
\begin{widetext}

\begin{eqnarray}\label{12}
    \hat{H}&=& 4E_{C_1}{\hat{n}_1}^{2}-E_{J_1}\cos{\hat{\phi}_1}+4E_{C_c}{\hat{n}_c}^{2} 
  -E_{J_c}\cos{\hat{\phi}_c}    +4E_{C_2}{\hat{n}_2}^{2}- E_{J_2}\cos{\hat{\phi}_2} \nonumber \\ &&  
  +8\frac{C_{1c}}{\sqrt{C_1C_c}}\sqrt{E_{C_1}E_{C_c}}({\hat{n}_1}{\hat{n}_c})  
   +8\frac{C_{2c}}{\sqrt{C_2C_c}}\sqrt{E_{C_2}E_{C_c}}({\hat{n}_2}{\hat{n}_c})    +8(1+\eta)\frac{C_{12}}{\sqrt{C_1C_2}}\sqrt{E_{C_1}E_{C_2}}({\hat{n}_1}{\hat{n}_2}),
    \end{eqnarray} 
    
\end{widetext}
where $\eta$ define as $(C_{1c}C_{2c})/(C_{12}C_c)$. We  assumed that the qubit-coupler capacitance   is  smaller than single-mode capacitance but larger than qubit-qubit coupling capacitance $C_{12}\ll C_{jc}\ll C_{\lambda}$. In the transmon regime, $E_{J_\lambda}/E_{C_\lambda}\gg 1$, the system Hamiltonian can be expressed as
\begin{equation}\label{13}
   \hat{H}=\hat{H}_1+\hat{H}_c+\hat{H}_2+\hat{H}_{1c}+\hat{H}_{2c}+\hat{H}_{12},
\end{equation}
\begin{equation}\label{14}
    \hat{H}=\omega_{\lambda}\hat{b}_{\lambda}^\dag \hat{b}_{\lambda}+\frac{\alpha_{\lambda}}{2}\hat{b}_{\lambda}^\dag\hat{b}_{\lambda}^\dag \hat{b}_{\lambda}\hat{b}_{\lambda}, \hspace{0.5cm}\lambda \in\{1,c,2\},
\end{equation}
\begin{equation}\label{15}
   \hat{ H}_{jc}=g_j(\hat{b}_{j}^\dag \hat{b}_{c}+\hat{b}_{j}\hat{b}_{c}^\dag -\hat{b}_{j}^\dag \hat{b}_{c}^\dag-\hat{b}_{j} \hat{b}_{c}),\hspace{0.5cm} j=1,2,
\end{equation}

\begin{equation}\label{16}
    \hat{H}_{12}=g_{12}(\hat{b}_{1}^\dag \hat{b}_2+\hat{b_1}\hat{b}_{2}^\dag -\hat{b}_{1}^\dag \hat{b}_{2}^\dag-\hat{b}_1 \hat{b}_2),
\end{equation}
where $\hat{b}_{\lambda}^\dag$ and $\hat{b}_{\lambda}$ are the creation and annihilation operators corresponding to each mode. And  $\omega_{\lambda}=\sqrt{8E_{J_\lambda}E_{C_\lambda}}-E_{J_\lambda}$ defines the oscillator frequency, $g_j=\frac{1}{2}\frac{C_{jc}}{\sqrt{C_{j}C_c}}\sqrt{\omega_{j}\omega_{c}}$  and $g_{12}=\frac{1}{2}(1+\eta)\frac{C_{12}}{\sqrt{C_{1}C_2}}\sqrt{\omega_{1}\omega_{2}}$ expresses qubit-coupler and qubit-qubit coupling strength respectively. In the dispersive regime, where, the coupler frequency is higher than the qubit frequency, the contributions from the double excitation and de-excitation will be significant that's why in Eq. (\ref{15}) we keep usual Jaynes-Cummings interaction term $(\hat{b}_{j}^\dag \hat{b}_c+\hat{b}_{j}\hat{b}_{c}^\dag)$ as well as counter-rotating term $(\hat{b}_{j}^\dag \hat{b}_{c}^\dag+\hat{b}_{j} \hat{b}_{c})$.

\begin{figure*}  
    \includegraphics[height=8cm,width=16cm]{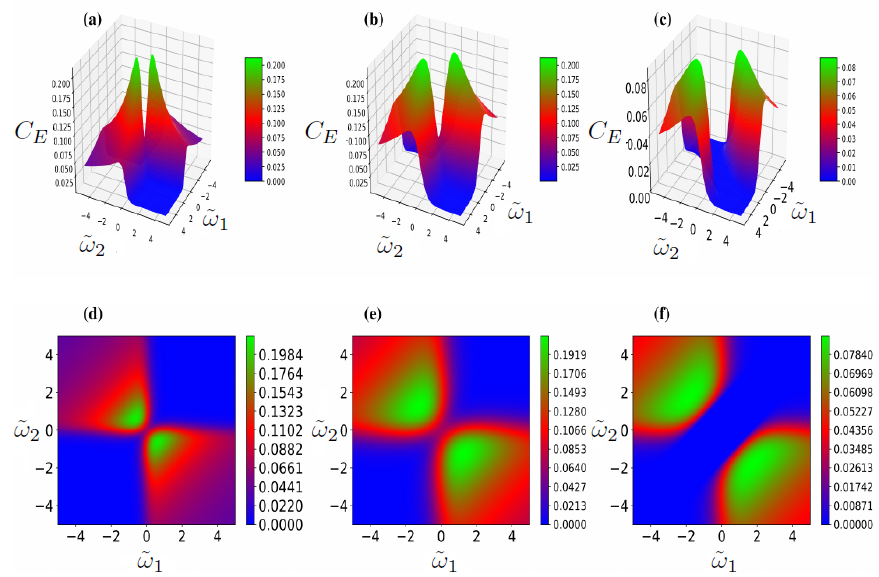}
    \caption{The two-dimensional concurrence in two  coupled qubit system is plotted versus dimensionless frequencies $\tilde{\omega}_1$ and $\tilde{\omega}_2$ for (a,d) $\tilde{g}=T=0.4$, (b,e) $\tilde{g}=T=0.2$  and (c,f) $\tilde{g}=0.2$ \&  $T=0.4$. $T$ is plotted in units of Boltzmann's constant where $k_B=1$.}
    \label{Figure03}
\end{figure*}

\section{Modeling and Schrieffer-Wolff Transformation}\label{Sec-3}
We consider a chain of three modes with exchange coupling between nearest neighbor (NN)  and next-nearest neighbors (NNN), as shown in Fig. \ref{Figure01}(b). The two qubits ($\omega_1$ and $\omega_2$) coupled to each other with coupling strength $g_{12}$, as well as each qubit couple to a tunable coupler ($\omega_c$) with coupling strength $g_i$ (i=1,2). Someone can achieve the two level effective Hamiltonian of our our proposed   system by changing   bosonic operators with Pauli operators as 
$b^{\dagger} \rightarrow \sigma^{+} ( b \rightarrow \sigma^{-} )$ and $b^{\dagger} b  \rightarrow (\ket{0}\bra{0} - \sigma_z)$.  Generally,   NN coupling is stronger than  NNN  coupling, i.e., $g_i>g_{12}>0$. The Hamiltonian for this two-level system is written as:
 
\begin{widetext}

\begin{equation}\label{24}
  {\hat{H}}=\sum_{j=1,2}\frac{1}{2}\omega_{j}\sigma_{j}^{z}+\frac{1}{2}\omega_{c}\sigma_{c}^{z}+\sum_{j=1,2}g_j(\sigma_{j}^{+}\sigma_{c}^{-}+\sigma_{j}^{-}\sigma_{c}^{+})+g_{12}(\sigma_{1}^{+}\sigma_{2}^{-}+\sigma_{1}^{-}\sigma_{2}^{+}),
\end{equation} 
\end{widetext}
where $\sigma_{\lambda}^{z}$, $\sigma_{\lambda}^{+}$ and $\sigma_{\lambda}^{-}$ ($\lambda= 1,2,c$) are Z, raising and lowering  Pauli operators, respectively. Here, $\omega_{1}$ and $\omega_{2}$ define the frequencies of qubit 1 and 2. These frequencies can be tuned by using energies of capacitors $E_{C_\lambda}$ and non-linear inductors $E_{J_\lambda}$. We define the detuning of two system as $\Delta_j\equiv\omega_j-\omega_c<0$, here $\omega_c$ represents the coupling circuit frequency. In Eq. (\ref{24}), $g_j$ characterize the coupling strength of qubit with coupler, and $g_{12}$ describes the coupling strength between the qubits. In our  proposed model, we use direct coupling and indirect coupling. The indirect coupling is modeled by using qubit, in literature it is called virtual-exchange interaction \cite{PhysRevA.75.032329}, that can be approximated by the Schrieffer-Wolff transformation (SWT) $U=exp[\sum_{j=1,2}(g_j/\Delta_j)(\sigma_{j}^{+}\sigma_{c}^{-}-\sigma_{j}^{-}\sigma_{c}^{+})]$ \cite{Bravyi-2011}. The SWT decouples the coupler from the system, resulting in effective two-qubit Hamiltonian for each mode,
\begin{equation}\label{25}
    \tilde{H}=\sum_{j=1,2}\frac{1}{2}\tilde{\omega}{_{j}}\sigma_{j}^{z}+\Big[\frac{g_1g_2}{\Delta}+g_{12}\Big](\sigma_{1}^{+}\sigma_{2}^{-}+\sigma_{1}^{-}\sigma_{2}^{+}),
\end{equation}
where, $\tilde{\omega}_j=\omega_j+{g_j}^{2}/\Delta_j$ defines the Lamb-shifted qubit frequency and the detuning expresses as $1/\Delta=(1/\Delta_1+1/\Delta_2)<0$. Also, we have considered that the coupler qubit stays in its ground state at all times. In Eq. (\ref{25}), the combined term inside the square brackets represents the total qubit-qubit effective coupling strength $\tilde{g}$, that can be controlled by the coupler frequency through $\Delta$, as well as by using  $g_1$ and $g_2$, both of which implicitly dependent on $\omega_c$. In general, $\tilde{g}$ is the function of the coupler and qubit frequencies. In the standard two-qubit basis $\ket{00}, \ket{01}, \ket{10}, \ket{11}$, the Hamiltonian of the system  can be expressed as 
\begin{equation}\label{26}
     \tilde{H} = \begin{pmatrix} 
   \frac{\alpha}{2}& 0 & 0 & 0\\
   0 & \frac{\tilde{\omega}}{2} & \tilde{g} & 0\\ 0 & \tilde{g} &-\frac{\tilde{\omega}}{2} & 0\\0 & 0 & 0 & -\frac{\alpha}{2}
    \end{pmatrix}.
\end{equation}
  The eigenvectors of Hamiltonian in Eq. (\ref{26}) are given as:
 
 \begin{equation*}
     \ket{\phi_1}=\ket{0,0},\hspace{0.5cm}\ket{\phi_2}=\ket{1,1},
 \end{equation*}
\begin{equation*}
     \ket{\phi_3}=\frac{1}{\sqrt{1+\frac{\xi^{2}}{4{\tilde{g}^{2}}}}}\Big(\frac{\xi}{2\tilde{g}}\ket{1,0}+\ket{0,1}\Big),
 \end{equation*}
 \begin{equation}\label{27}
     \ket{\phi_4}=\frac{1}{\sqrt{1+\frac{\zeta^{2}}{4{\tilde{g}^{2}}}}}\Big(\frac{\zeta}{2\tilde{g}}\ket{1,0}+\ket{0,1}\Big),
 \end{equation}
with corresponding eigen energies: $ E_1=-\frac{\alpha}{2},$ $E_2=\frac{\alpha}{2},$ $ E_3=-\frac{\gamma}{2},$ and $E_4=\frac{\gamma}{2},$ where $\alpha=\tilde{\omega}_1+\tilde{\omega}_2$, $\tilde{\omega}=\tilde{\omega}_1-\tilde{\omega}_2$, $\gamma=\sqrt{4\tilde{g}^{2}+\tilde{\omega}^{2}}$, $\xi=-\tilde{\omega}+\gamma$ and $\zeta=-(\tilde{\omega}+\gamma)$. Note that when $\tilde{\omega}\rightarrow0$ and $\tilde{g}>0$, the two states   go to the maximally entangled states as  $\ket{\phi_3}=\frac{1}{\sqrt{2}}(\ket{01}+\ket{10})$ and $\ket{\phi_4}=\frac{1}{\sqrt{2}}(\ket{01}-\ket{10})$. For $\tilde{g}<0$, both states are converted to $ \ket{\phi_3}=\frac{1}{\sqrt{2}}(\ket{01}-\ket{10})$ and $\ket{\phi_4}=\frac{1}{\sqrt{2}}(\ket{01}+\ket{10})$. Moreover, we can find that the eigen energies are even function of the effective qubit-qubit coupling $\tilde{g}$. Therefore,  the ground-state entanglement exists for both positive direct coupling and negative indirect coupling and it should be symmetric with respect to the effective qubit-qubit coupling $\tilde{g}$. The ground state entanglement depends on the effective qubit-qubit coupling constant $\tilde{g}$, $\alpha$ and $\gamma$, as predicted in Eq. (\ref{27}).

 \begin{figure}  
    \includegraphics[height=8cm,width=9cm]{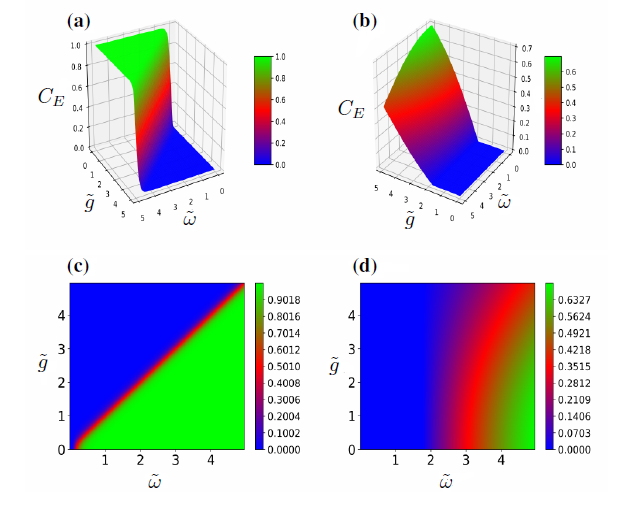}
    \caption{The two-dimensional concurrence in two  coupled qubit system is plotted versus $\tilde{g}$ and $\tilde{\omega}$ for different values of dimensionless temperature (a,c) $T=0.1$ and for (b,d) $T=2$.}
    \label{Figure04}
\end{figure}

\subsection{Thermal Entanglement}
 In this subsection, we highlight how quantum entanglement depends on external temperature by making use of the concurrence. The thermal density matrix $\rho(T)$ for this system in terms of standard basis can be written as 
\begin{equation}\label{28}
    \rho(T)=\frac{e^{-\beta H}}{Z}=\frac{1}{Z}\sum_{i=1}^{4}e^{-\beta E_{m}}\ket{\psi_i}\bra{\psi_i},
\end{equation}
\begin{equation}\label{29}
  \rho(T)=  \frac{1}{Z}\begin{pmatrix} 
  \rho_{11}&  0 & 0 & 0\\
   0 & \rho_{22} & \rho_{23} & 0\\ 0 & \rho_{32} & \rho_{33}& 0\\0 & 0 & 0 & \rho_{44}
    \end{pmatrix},
\end{equation}
where $\beta=1 /( k_{B}T)$, $\rho_{11}=e^{-{\alpha}/{2T}}$,    $\rho_{22}=\rho_{33}=\frac{\gamma^{2}\cosh\big({\frac{\gamma}{2T}}\big)+\gamma\tilde{\omega}\sinh\big({\frac{\gamma}{2T}}\big)}{\gamma^{2}}$, $\rho_{23}=\rho_{32}=\frac{2\tilde{g}\sinh\big({\frac{\gamma}{2T}}\big)}{\gamma}$ and $\rho_{44}=e^{{\alpha}/{2T}}$. For convenience, we choose $k_{B}=1$ throughout these calculations.   However, someone can get SI units of temperature by using this relation  $T' = \frac{\hbar  \omega }{k_B T} $, here $\omega  \sim 4 $ GHz and $T' $ is defined in mK.   
The partition  function $Z$ is calculated as
\begin{equation}\label{30}
    Z=2\Big[\cosh\Big({\frac{\alpha}{2T}}\Big)+\cosh\Big({\frac{\gamma}{2T}}\Big)\Big].
\end{equation}
 The concurrence $0$ corresponds to separable or unentangled states and $1$ defines maximally entangled states and is expressed as  \cite{PhysRevA.67.012307},
\begin{equation}\label{31}
    C_E=max\Big[0,\lambda_{1}-\lambda_{2}-\lambda_{3}-\lambda_{4}\Big],
\end{equation}
where, $\lambda_{i}$'s are the square root of the eigenvalues of the matrix $R$, which we can calculate with the help of Eq. (\ref{29}). To observe the effect of coupling strength, temperature, and qubit frequencies on the quantum entanglement, we divide our discussion into three special cases.

\textbf{Case 1:} In the first scenario, we would like to discuss the dependence of quantum entanglement on the coupling strength $\tilde{g}$ and on the temperature $T$ as shown in Fig. \ref{Figure02}. In this special case, we let the qubit frequencies are constant. We let $\tilde{\omega}_1=\tilde{\omega}_2=1$, we notice that the two qubits are maximally entangled for $\tilde{g} = 5$ and for $T \rightarrow 0$. However, as the temperature started to increase the concurrence is started to become zero. It means that the two qubits are less entangled for high temperature, as shown in Fig. \ref{Figure02}(a). On the other hand, if  $\tilde{\omega}_1 \neq \tilde{\omega}_2$, the entanglement gets vanishes for low temperature too, even though the coupling strength is quite high as shown in Fig. \ref{Figure02}(b). Additionally, for higher coupling strength, the entanglement strength is quite weak as compared to the first scenario, when qubit frequencies are equal. Therefore, our results suggest that for strong entanglement it is better to have superconducting qubits with the same frequencies.

\textbf{Case 2:} To see the effect of superconducting qubit frequency on the quantum entanglement for constant temperature and coupling strength, we plot Fig. \ref{Figure03}. The values of the frequencies can be tuned by using $E_J$  and $E_C$. We can achieve the negative regime of the frequency by letting $\frac{E_J}{E_C}<\frac{1}{8}$. We notice that the quantum entanglement vanishes when both frequencies become zero, which is indeed true as there is no qubit left for the entanglement. We also observe that for small values of equal frequencies, the chances of qubits entanglement are high as compare to high frequencies. On the other hand, if the two-qubit frequencies are different from each other, then there are very less chances that qubits get entanglement. If one qubit frequency is negative and the other is positive then we find high chances to get quantum entanglement as shown in Fig. \ref{Figure03}(b), for $\tilde{g}=T=0.2$. For coupling strength $\tilde{g}=0.2$ and temperature $T=0.4$, we notice that the entanglement is only possible for high-frequency superconducting qubits. In another way, we can say that it is less probable that the two qubits get entangled for lower frequencies, high temperature, and low coupling strength.

\textbf{Case 3:}  In this special case, we calculate quantum entanglement dependence on the coupling strength and frequencies of the qubits, for a constant temperature. We note that for $T=0.1$, the concurrence is 1 for small values of coupling strength and for high values of frequencies as shown in Fig. \ref{Figure04}(a) and \ref{Figure04}(c). However, as we increase the frequency the concurrence goes to zero for low coupling strength, as shown in \ref{Figure04}(c). However, as we increase the temperature $T=2$, the transition from entangle states to unentangle states happen near $\tilde{\omega} \leq 2$. This graph shows that the entangled states are possible for higher frequencies and low coupling strength.

\section{Conclusion}\label{Sec-4}
 In conclusion, by using the concept of concurrence we have investigated the thermal entanglement in two coupled superconducting qubits under the influence of arbitrary coupling strength. Our proposed model is quite realistic, therefore, it can be experimentally implemented and all the parameters are tunable. We have found that the thermal entanglement can be efficiently controlled through the effective qubit-qubit coupling strength, qubit frequencies, and temperature. The effects of affective qubit-qubit coupling strength and temperature on entanglement are studied for different values of qubit frequencies. We find that entanglement exists and it can be enhanced by using the coupled qubits with the same frequencies. Our results imply that the two coupled qubits can go to maximally entangled states at low temperature and for equal and high values of qubit frequencies. Our proposed model can further be enhanced for gates, where, someone can study the effect of temperature on the fidelity of quantum gates. This model can also be beneficial for study thermal entanglement for many ($n>2$) qubit states. Our findings identify the real potential to study thermal entanglement, especially for superconducting qubits.

\section{Acknowledgment} \label{Sec-6}

Javed Akram thank Hamza Qayyum, Tasawar Abbas, Sajid Alvi and Waqas Masood for the insightful discussion.  
 
% 
% \section{Appendix A}
% 
% Complete solution for the  Mean position of the BEC including  ''non-conservative`` part of the Lagrangian 
%  

%\bibliographystyle{acm}

\section*{References}

%\bibliography{Ent-bibliography}

\end{document}